\documentclass[%
reprint,
 preprintnumbers,
 amssymb,
 aps,
]{revtex4-1}

\usepackage{amsmath}
\usepackage[dvipdfmx]{graphicx}
\usepackage{dcolumn}
\usepackage{bm}
\usepackage{braket}

\usepackage{ulem}
\usepackage{color}
\usepackage{cleveref} 

\newcommand{\Slash}[1]{{\ooalign{\hfil/\hfil\crcr$#1$}}}

\begin{document}
\preprint{}

\title{
$\mu^- \to e^-\gamma$ in a muonic atom as a probe for effective lepton flavor violating operators involving photon fields
}

\author{Yuichi Uesaka$^{1}$,
Masato Yamanaka$^{2}$,
and Yoshitaka Kuno$^{3,4}$}
\affiliation{
$^1$Department of Fundamental Education, Dokkyo Medical University, 880 Kitakobayashi, Mibu,
Shimotsuga, Tochigi 321-0293, Japan\\
$^2$Department of Advanced Sciences, Hosei University, Tokyo 184-8584, Japan\\
$^3$Department of Physics, Osaka University, Toyonaka, Osaka 560-0043, Japan\\
$^4$Research Center of Nuclear Physics, Osaka University, Osaka 567-0047, Japan\\
}




\date{\today}

\begin{abstract}
We propose the $\mu^-\to e^-\gamma$ process in a muonic atom as a novel means to investigate charged lepton flavor violation (CLFV). We demonstrate its sensitivity in probing effective CLFV operators associated with both single and double photon fields.
In comparison to $\mu^+\to e^+\gamma$ using free positive muon decays at rest, the emitted electron and photon in $\mu^-\to e^-\gamma$ exhibit non-monochromatic spectra, presenting non-trivial case.
We derive the decay rate formula and demonstrate that the potential rate of the process is significant enough to motivate exploration in future muon CLFV experiments.
\end{abstract}

\maketitle

\onecolumngrid


\section{Introduction \label{sec:Intro}}

Charged lepton flavor violation (CLFV) is known to serve as a crucial probe for searching for new physics beyond the Standard Model (SM)~\cite{Kuno:1999jp,Calibbi:2017uvl}.
With various potential sources to contribute to CLFV, it is important to explore diverse CLFV processes to obtain insights into new physics.
Muon rare decays become a powerful avenue for investigating CLFV processes characterized with $\Delta L_\mu=-\Delta L_e=\pm 1$.
High intensity muon beams available today can produce $\mathcal{O}\left(10^{\textcolor{blue}{7-}8}\right)$ muons per second, yielding the current upper limits of branching ratios of typically $\mathcal{O}\left(10^{-13}\right)$, as shown in Table~\ref{tab:experimental_bounds}.
In upcoming future, several experiments will improve higher sensitivity to update these constraints by several orders of magnitude.
\begin{table}[htb]
\centering
 \begin{tabular}{l|ll|cl}
 Process & Current bound & & Future sensitivity & \\
 \hline\hline
$\mu^+\to e^+\gamma$ & $ < 3.1 \times 10^{-13}$ & MEG II \& MEG~\cite{MEGII:2023ltw} & $\mathcal{O}\left(10^{-14}\right)$ & MEG II~\cite{MEGII} \\
 $\mu^+\to e^+\gamma\gamma$ & $ < 7.2 \times 10^{-11}$ & Crystal Box~\cite{CrystalBox} & None &  \\
 $\mu^+\to e^+e^+e^-$& $ < 1.0  \times 10^{-12}$ & SINDRUM~\cite{Bellgardt:1987du} & $  \mathcal{O}\left(10^{-16}\right)$ & Mu3e~\cite{mu3e} \\
 $\mu^-N\to e^-N$ & $< 7 \times 10^{-13}$ & SINDRUM II~\cite{Bertl:2006up} & $\mathcal{O}\left(10^{-16}\right)$ & COMET~\cite{COMET2018}, Mu2e~\cite{Bartoszek2015} \\
  & $< 6.1 \times 10^{-13}$ & SINDRUM II \cite{Wintz:1998rp} & $\mathcal{O}\left(10^{-18}\right)$ & PRISM/PRIME~\cite{PP} \\ \hline
\end{tabular}
 \caption{Current experimental bounds (90\% confidence level) on CLFV muon rare decay branching ratios, along with projected sensitivities for future experiments.
 The 5th and 6th rows show muon-to-electron conversion in muonic atom, $\mu^-N\to e^-N$, where $N$ represents a nucleus.
 The current bounds from SINDRUM II~\cite{Bertl:2006up} and SINDRUM II~\cite{Wintz:1998rp} are given in $N=$Au (gold) and $N=$Ti (titanium), respectively.
 In the future COMET~\cite{COMET2018} and Mu2e~\cite{Bartoszek2015} experiments, $N=$Al (aluminum) will be used.
   \label{tab:experimental_bounds} }
 \end{table}

The effective field theory (EFT) approach is valuable for studying rare processes in model-independent way. It involves possible effective operators allowed by the symmetries of the SM.
One of the interesting CLFV operator is the diphoton CLFV operators like $\overline{e}\mu F_{\alpha\beta}F^{\alpha\beta}$.
These effective operators are often referred to as \textit{Rayleigh operators} because they resemble the operators associated with Rayleigh scattering.
The effective diphoton operators have a higher mass dimension than the dipole operators $\overline{e}\sigma_{\alpha\beta}\mu F^{\alpha\beta}$, which might suggest they are less significant in the naive sense of EFT.
However, several models are known to produce a higher rate of $\mu\to e\gamma\gamma$ compared to $\mu\to e\gamma$; e.g., models including doubly-charged leptons~\cite{Wilczek:1977wb,Bowman:1978kz}, vector leptoquarks~\cite{Gvozdev:1994bk}, scalars with off-diagonal Yukawa interaction to charged leptons~\cite{Gemintern:2003gd,Cordero-Cid:2005vca}, and excited and fourth generation leptons~\cite{Inan:2011gb}.
The diphoton operators are directly constrained by the three-body rare decay of $\mu^+\to e^+\gamma\gamma$; the upper limit of the branching ratio is given by $BR\left(\mu^+\to e^+\gamma\gamma\right)<7.2\times 10^{-11}$~\cite{CrystalBox}.
However, as shown in \Cref {tab:experimental_bounds}, no future experimental attempts have been planned to directly measure the process $\mu^+\to e^+\gamma\gamma$.
Also as summarized in Ref.~\cite{Fortuna:2023paj}, it has been pointed out that the diphoton operators can also be investigated indirectly by $\mu^+\to e^+\gamma$~\cite{Fortuna:2022sxt} and $\mu^-\to e^-$ conversion~\cite{Davidson:2020ord,Haxton:2024lyc}.

\begin{figure}[htb]
  \centering
  \includegraphics[clip, width=5.0cm]{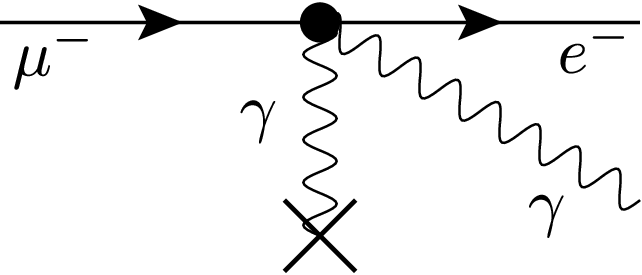}
  \caption{
  Diagram of $\mu^-\to e^-\gamma$ in a muonic atom via the CLFV diphoton operator, described by Eq.~\eqref{eq:Lagrangian_diphoton}.
  The initial muon is in an atomic orbit, and there are an electron and a photon in the final state.
  The blob represents the vertex of the effective CLFV interaction and the cross represents the nuclear Coulomb field.
  }
  \label{fig:diagram}
\end{figure}

In this article, as a new strategy to probe the diphoton CLFV operators, we consider $\mu^-\to e^-\gamma$ process in a muonic atom.
Since the nucleus can absorb one photon from the CLFV vertex, the diphoton operators contribute as shown by Fig.~\ref{fig:diagram}.
Since this kind of the process can occur in the presence of a nucleus, the direct search cannot be performed by positive muons.
The experimental signature of the $\mu^-\to e^-\gamma$  decay is a pair of an electron and a single photon with the specific energy sum, $E_{e^-}+E_{\gamma}=m_\mu-B$, where $B$ is the binding energy of the initial muon in the muonic atom.

The dipole CLFV operators induce the direct decay of a muon in orbit, $\mu^{-}\to e^{-}\gamma$, producing the same final states as the diphoton operators.
The difference in mechanisms between the diphoton and dipole operators is evident in the momentum distributions of the emitted particles observed in experimental searches, making it possible to identify the dominant operator. Our discussion includes contributions from both diphoton and dipole operators.

This article is organized as follows:
In Sec.~\ref{sec:Formalism}, we derive the formulas for the decay rate of $\mu^-\to e^-\gamma$ 
process in a muonic atom via the effective diphoton and dipole CLFV operators, taking into account the bound-state effects of the initial muon.
Based on the obtained formulas,  in Sec.~\ref{sec:Results}, we present the results of the numerical computations and discuss the potential of search for the process, noting that the current constraints suggest the $\mu^-\to e^-\gamma$ rate in a muonic atom could be sizable.
Background estimation is discussed in Sec.~\ref{sec:Background}.
Finally, we give a summary in Sec.~\ref{sec:Summary}.

\section{Formulation for $\mu^-\to e^-\gamma$ in muonic atoms \label{sec:Formalism}}

We introduce the effective CLFV interaction Lagrangian as follows:
\begin{align}
\mathcal{L}_\mathrm{eff}=\mathcal{L}_D+\mathcal{L}_{FF},
\label{eq:Lagrangian}
\end{align}
with the dipole operators,
\begin{align}
\mathcal{L}_{D}=&\frac{m_\mu}{v^2}\overline{e}\sigma_{\alpha\beta}\left(D_{L}P_L+D_{R}P_R\right)\mu F^{\alpha\beta}+[H.c.],
\label{eq:Lagrangian_dipole}
\end{align}
and the diphoton ones,
\begin{align}
\mathcal{L}_{FF}=&\frac{1}{v^3}\left[\overline{e}\left(C_{L}P_L+C_{R}P_R\right)\mu F_{\alpha\beta}F^{\alpha\beta}+\overline{e}i\gamma_5\left(\tilde{C}_{L}P_L+\tilde{C}_{R}P_R\right)\mu F_{\alpha\beta}\tilde{F}^{\alpha\beta}\right]+[H.c.].
\label{eq:Lagrangian_diphoton}
\end{align}
Here, $P_{L/R}=\left(1\mp\gamma_5\right)/2$ are the chirality projection operators, $F_{\alpha\beta}=\partial_\alpha A_\beta-\partial_\beta A_\alpha$ is the 
strength of the photon field $A_\alpha$, and $\tilde{F}^{\alpha\beta}=\epsilon^{\alpha\beta\gamma\delta}F_{\gamma\delta}/2$ is its dual tensor.
The dimensionless coupling constants are given as $D_L$, $D_R$, $C_L$, $C_R$, $\tilde{C}_L$, and $\tilde{C}_R$, and the overall normalization is defined by $v=\sqrt{1/(2\sqrt{2}G_F)}$ with the Fermi constant $G_F=1.166\times 10^{-11}$\,MeV$^{-2}$.

We consider that the initial muon is in the $1S$ bound state within the nuclear Coulomb field, and the final state consists of the pair of an electron and a single photon.
The electron and photon are mostly emitted in back-to-back, and the typical momentum transfer $p_N$ to the nucleus is around $Z\alpha m_\mu$.
Then, the energy transfer $p_N^2/\left(2m_N\right)$, where $m_N$ is the nuclear mass, is estimated to be less than a few keV for $Z\simeq 10$, which is negligibly small.
Ignoring the nuclear recoil energy, the rate of $\mu^-\to e^-\gamma$ in a muonic atom is given as
\begin{align}
d\Gamma=&\frac{d^3p_e}{(2\pi)^32E_e}\frac{d^3p_\gamma}{(2\pi)^32E_\gamma}(2\pi)\delta\left(E_e+E_\gamma-E_\mu\right)\frac{1}{2}\sum_{spins}\left|\mathcal{M}\right|^2,
\end{align}
where the delta function ensures the energy conservation with the energy $E_i$ of the particle $i=e,\gamma,\mu$; it is satisfied that $E_e+E_\gamma=E_\mu$.
The amplitude $\mathcal{M}$ is
\begin{align}
\mathcal{M}=&-\frac{2im_\mu}{v^2}\int d^3r\overline{\psi}_e\left(\bm{r}\right)\sigma_{\alpha\beta}\left(D_{L}P_L+D_{R}P_R\right)\psi_\mu^{1S}\left(\bm{r}\right)p_\gamma^\alpha\epsilon^{s_\gamma *\beta}\exp\left(-i\bm{p}_\gamma\cdot\bm{r}\right) \nonumber\\
&-\frac{4i}{v^3}\int d^3r\overline{\psi}_e\left(\bm{r}\right)\left(C_{L}P_L+C_{R}P_R\right)\psi_\mu^{1S}\left(\bm{r}\right)p_\gamma^\alpha\epsilon^{s_\gamma *\beta}\exp\left(-i\bm{p}_\gamma\cdot\bm{r}\right)\Braket{N|F_{\alpha\beta}|N} \nonumber\\
&-\frac{4i}{v^3}\int d^3r\overline{\psi}_e\left(\bm{r}\right)i\gamma_5\left(\tilde{C}_{L}P_L+\tilde{C}_{R}P_R\right)\psi_\mu^{1S}\left(\bm{r}\right)p_\gamma^\alpha\epsilon^{s_\gamma *\beta}\exp\left(-i\bm{p}_\gamma\cdot\bm{r}\right)\Braket{N|\tilde{F}_{\alpha\beta}|N}.
\end{align}
The expectation values of the field strength in a nucleus can be replaced with a electric field $\bm{E}$ as
\begin{align}
\Braket{N|F_{\alpha\beta}|N}=
\begin{cases}
-E_i & (\alpha=i,\beta=0) \\
E_j & (\alpha=0,\beta=j) \\
0 & (\alpha=i,\beta=j)
\end{cases},
\end{align}
and
\begin{align}
\Braket{N|\tilde{F}_{\alpha\beta}|N}=
\begin{cases}
-\epsilon_{ijk}E_k & (\alpha=i,\beta=j) \\
0 & (\alpha=0\text{ or }\beta=0)
\end{cases}.
\end{align}

The lepton wave functions, $\psi_\mu^{1S}$ and $\psi_e$, are the four component spinor satisfying the Dirac equation in the nuclear Coulomb potential,
\begin{align}
    \left[i\Slash{\partial}-m_\ell+V(\bm{r})\gamma_0\right]\psi_\ell\left(\bm{r}\right)=0.
\end{align}
The nuclear Coulomb potential $V(r)$ is determined by the nuclear charge density $\rho(\textbf{r})$,
\begin{align}
    V(\bm{r})=&\frac{e}{4\pi}\int d\bm{r}'\frac{\rho\left(\bm{r}'\right)}{\left|\bm{r}'-\bm{r}\right|}.
\end{align}

In this article, we employ the following approximation, which is valid for light atoms:
First, we ignore the nuclear Coulomb potential when calculating the electron wave functions and instead use a plane wave approximation.
\begin{align}
\overline{\psi}_e\left(\bm{r}\right)=& \overline{u_e^{s_e}}\left(p_e\right)\exp\left(-i\bm{p}_e\cdot\bm{r}\right).
\end{align}
Second, we take a point-like distribution of the nuclear charge density,
\begin{align}
\rho\left(\bm{r}\right)=Ze\delta^{(3)}\left(\bm{r}\right).
\end{align}
The point-like nucleus makes the electric field $E(r)=Ze/(4\pi r^2)$, where $r$ is the distance from the nuclear center.
Third, for the bound muons, we use the non-relativistic wave function in the point-like nuclear Coulomb potential,
\begin{align}
\psi_\mu^{1S}\left(\bm{r}\right)=&\sqrt{\frac{\left(m_\mu\zeta\right)^3}{\pi}}\exp\left(-m_\mu\zeta r\right)
\begin{pmatrix}
\chi_{1/2}^{s_\mu} \\
0
\end{pmatrix}.
\end{align}
Here, we define $\zeta=Z\alpha$, where $Z$ is the atomic number of the nucleus and $\alpha=e^2/(4\pi)$ is the QED fine structure constant.
The energy of the bound muon is given by $E_\mu=\left(1-b\right)m_\mu$, where $bm_\mu$ is the binding energy of the $1S$ muon and $b=\zeta^2/2$ is obtained under the non-relativistic approximation for the point-like nucleus.

Hereafter we use the dimensionless energies, $x=E_e/m_\mu$ and $y=E_\gamma/m_\mu$.
We also define $r=\left(m_e/m_\mu\right)^2$.
In the signal event, the energy conservation ensures the relation that $x+y=1-b$, and the allowed ranges of $x$ and $y$ are respectively given as $\sqrt{r}<x<1-b$ and $0<y<1-b-\sqrt{r}$.

Under the low-$Z$ approximation, we analytically obtain the energy and angular distribution,
\begin{align}
\frac{1}{\Gamma_0}\frac{d\Gamma\left(\mu^-\to e^-\gamma\right)}{dxd\cos\theta_{e\gamma}}=& 3072\pi\zeta^5\sqrt{x^2-r}y^3\left[f_D^2xd\left(\left|D_L\right|^2+\left|D_R\right|^2\right)\right. \nonumber\\
&+f_Df_C\sqrt{x^2-r}\sin^2\theta_{e\gamma}\mathrm{Re}\left[D_LC_L^*+D_RC_R^*+D_L\tilde{C}_L^*+D_R\tilde{C}_R^*\right] \nonumber\\
&\left.+f_C^2\sin^2\theta_{e\gamma}\left\{x\left(\left|C_L\right|^2+\left|C_R\right|^2+\left|\tilde{C}_L\right|^2+\left|\tilde{C}_R\right|^2\right)+2\sqrt{r}\mathrm{Re}\left[C_LC_R^*-\tilde{C}_L\tilde{C}_R^*\right]\right\}\right],
\label{eq:ene_ang_distribution}
\end{align}
where $y=1-b-x$, and the normalization is given by the free muon decay rate $\Gamma_0=m_\mu^5/\left(1536\pi^3v^4\right)$.
Here, we use the following abbreviations:
\begin{align}
f_D=&\frac{4}{\left(w^2+\zeta^2\right)^2}, \\
f_C=&\frac{1}{\sqrt{\pi\alpha}}\frac{m_\mu}{v}\frac{\sqrt{x^2-r}}{w^2}\left\{1-\frac{\zeta}{w}\tan^{-1}\left(\frac{w}{\zeta}\right)\right\},
\end{align}
where $w=\sqrt{x^2-r+y^2+2xy(1-d)}$.
The angle $\theta_{e\gamma}$ is defined as the opening angle between the momenta of the emitted electron and photon, and
\begin{align}
     d=1-\frac{\sqrt{x^2-r}}{x}\cos\theta_{e\gamma}.
\end{align}

\section{Numerical analyses for the signal process \label{sec:Results}}

By numerically calculating Eq.~\eqref{eq:ene_ang_distribution}, we can discuss the dependence on the atomic number ($Z$) and the kinematic distributions of the emitted particles.
Before proceeding with the discussion, we will first examine the constraints for the coupling constants.

The magnitudes of the coupling constants are constrained by experiments searching for rare decays of free muons.
The upper limit on the couplings of the dipole operators is determined by the $\mu^+\to e^+\gamma$ search; the branching ratio is given by
\begin{align}
BR\left(\mu^+\to e^+\gamma\right)=384\pi^2\left(\left|D_L\right|^2+\left|D_R\right|^2\right),
\label{eq:br_meg}
\end{align}
where the electron mass is ignored.
We note that Eq.~\eqref{eq:br_meg} is consistently obtained by investigating the $Z\to 0$ limit of Eq.~\eqref{eq:ene_ang_distribution}.
The most stringent bound of the branching ratios is experimentally given as $BR(\mu^+\to e^+\gamma)<3.1\times 10^{-13}$ by MEG II~\cite{MEGII:2023ltw}.
Using Eq.~\eqref{eq:br_meg}, it is translated to the upper limit of the dipole coupling constants, $\sqrt{\left|D_L\right|^2+\left|D_R\right|^2}<9.0\times 10^{-9}$.

On the other hand, the couplings of the diphoton operators are directly constrained by the search for the $\mu^+\to e^+\gamma\gamma$ process.
Ignoring the electron mass, we find the branching ratio of $\mu^+\to e^+\gamma\gamma$,
\begin{align}
BR\left(\mu^+\to e^+\gamma\gamma\right)=&\frac{2}{5}\left(\frac{m_\mu}{v}\right)^2\left(\left|C_L\right|^2+\left|C_R\right|^2+\left|\tilde{C}_L\right|^2+\left|\tilde{C}_R\right|^2\right).
\label{eq:br_megg}
\end{align}
The current experimental constraint, provided by the Crystal Box experiment~\cite{CrystalBox}, is  $BR(\mu^+\to e^+\gamma\gamma)<7.2\times 10^{-11}$.
According to Eq.~\eqref{eq:br_megg}, this constraint corresponds to $\sqrt{|C_L|^2+|C_R|^2+|\tilde{C}_L|^2+|\tilde{C}_R|^2}<2.2\times 10^{-2}$.

Another constraint on $C_L$ and $C_R$ comes from the search for $\mu^-\to e^-$ conversion~\cite{Davidson:2020ord}.
Ignoring other effective operators, we obtain $\sqrt{|C_L|^2+|C_R|^2}<1.0 \times 10^{-3}$ from the result of the SINDRUM II experiment, $BR(\mu^-\textrm{Au}\to e^-\textrm{Au})<7\times 10^{-13}$~\cite{Bertl:2006up}.
This naively provides a stronger constraint on $C_L$ and $C_R$ than the Crystal Box experiment.
However, since the various effective operators (even those involving quark fields or gluon fields) generally interfere, there are still degrees of freedom to cancel the $\mu^-\to e^-$ conversion rate.
The constraint from the $\mu^-\to e^-$ conversion is complementary to the one from $\mu^+\to e^+\gamma\gamma$, which directly restricts $\mu^-\to e^-\gamma$ in muonic atoms.

Since the other diphoton couplings $\tilde{C}_L$ and $\tilde{C}_R$ do not contribute to the \textit{coherent} $\mu^-\to e^-$ conversion, their constraints from the $\mu^-\to e^-$ searches are not stringent.
The quantitative estimation for the diphoton operators including $F\tilde{F}$ are recently discussed in Ref.~\cite{Haxton:2024lyc}.

Also, the diphoton couplings are indirectly constrained by $\mu^+\to e^+\gamma$~\cite{Fortuna:2022sxt}.
The branching ratio of $\mu^+\to e^+\gamma$ is estimated as
\begin{align}
BR\left(\mu^+\to e^+\gamma\right)\simeq\frac{6\alpha}{\pi}\left(\frac{m_\mu}{v}\right)^2\log^2\left(\frac{\Lambda^2}{m_\mu^2}\right)\left(\left|C_L+\tilde{C}_L\right|^2+\left|C_R+\tilde{C}_R\right|^2\right),
\end{align}
where $\Lambda$ is the cutoff scale of the effective theory.
For instance, setting $\Lambda=100$\,GeV, we obtain an indirect constraint for the diphoton couplings, $\sqrt{|C_L+\tilde{C}_L|^2+|C_R+\tilde{C}_R|^2}<5.7\times 10^{-4}$, which is stronger than the direct constraint from $\mu^+\to e^+\gamma\gamma$.
We note that the $C$ and $\tilde{C}$ terms generally interfere, although the interference is neglected in Ref.~\cite{Fortuna:2022sxt}.
Considering the interference, we find that the $\mu^+\to e^+\gamma$ rate disappears in case of $C_{L/R}=-\tilde{C}_{L/R}$ while the $\mu^+\to e^+\gamma\gamma$ rate cannot be canceled.
In the latter discussion, we use the bound from the Crystal Box experiment as the constraint for the diphoton couplings.

For simplicity, throughout the rest of this section, we use only $D_L$ and $C_L'\equiv\sqrt{|C_L|^2+|\tilde{C}_L|^2}$, setting all other coupling constants, namely $D_R$, $C_R$, and $\tilde{C}_R$, to zero.

\subsection{Atomic number dependence}

\begin{figure}[htb]
  \centering
    \begin{tabular}{c}
      \begin{minipage}{0.45\hsize}
		\centering
          \includegraphics[clip, width=7.0cm]{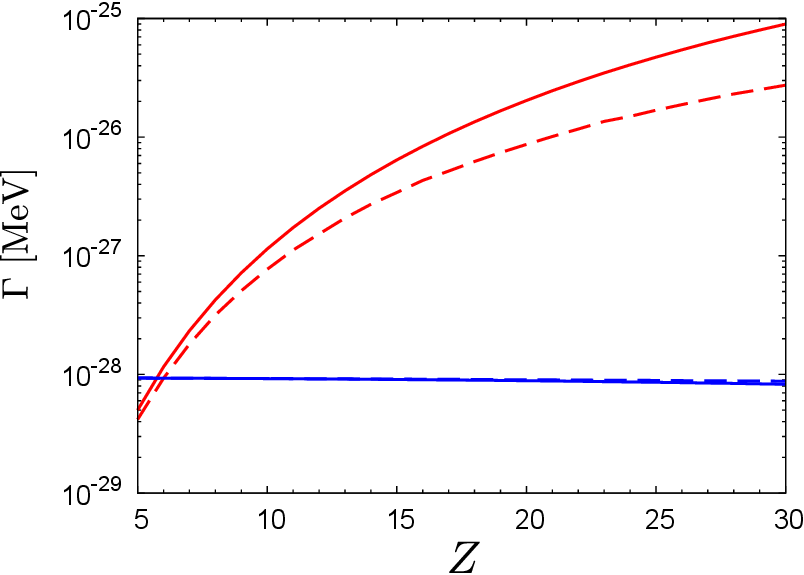}
          \\ (a)
      \end{minipage}%
      \begin{minipage}{0.45\hsize}
        \centering
          \includegraphics[clip, width=7.0cm]{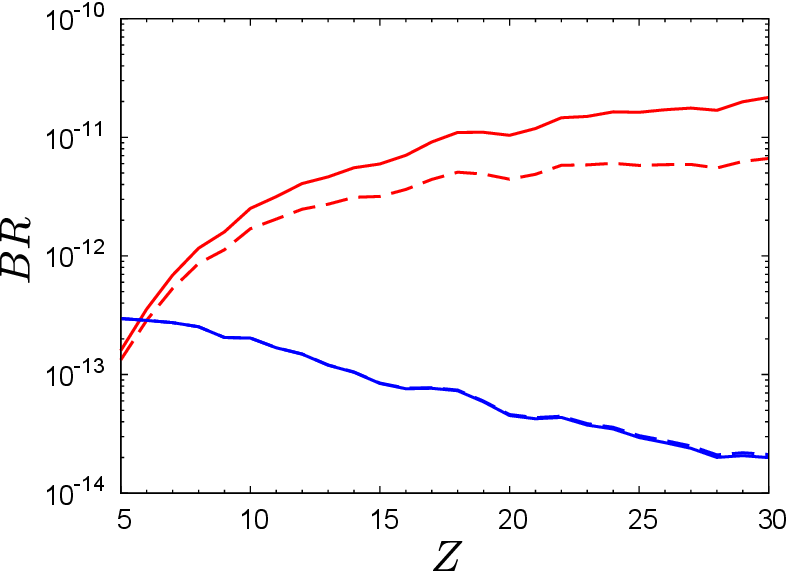}
          \\ (b)
      \end{minipage}%
    \end{tabular}
    \caption{
    (a) The $Z$-dependence of the $\mu^-\to e^-\gamma$ rate and (b) the branching ratios.
    The horizontal axis is the atomic number $Z$ of the target nucleus.
   	Red curves indicate the case of $C_L'=2.2\times 10^{-2}$ and $D_L=D_R=C_R=\tilde{C}_R=0$ are set.
   	Blue curves indicate the case of $D_L=1.0\times 10^{-8}$ and $C_L'=D_R=C_R=\tilde{C}_R=0$ are set.
   	On the solid and dashed curves, $Z$ and $Z_\mathrm{eff}$ are respectively used in the calculation.
    }
    \label{fig:rate_vs_z_crystalbox}
\end{figure}

We consider the $Z$ dependence of the decay rate calculated by integrating Eq.~\eqref{eq:ene_ang_distribution} by $\sqrt{r}\le x\le 1-b$ and $-1\le\cos\theta_{e\gamma}\le 1$ for various nuclei with the atomic number $Z$.
The allowed maximum values of the rate for various $Z$ are shown by the solid curves in Fig.~\ref{fig:rate_vs_z_crystalbox}-(a), and the corresponding branching ratio is shown in Fig.~\ref{fig:rate_vs_z_crystalbox}-(b).
We plot the values only for nuclei with $Z\le 30$ because the accuracy of our estimation would be insufficient for heavy nuclei.
Here, the branching ratio is calculated by
\begin{align}
    BR(\mu^-\to e^-\gamma)=\tilde{\tau}_\mu\int_{\sqrt{r}}^{1-b}dx\int_{-1}^{1}d\cos\theta_{e\gamma}\frac{d\Gamma}{dxd\cos\theta_{e\gamma}},
\end{align}
where $\tilde{\tau}_\mu$ is the lifetime of a muonic atom.
The experimental values of $\tilde{\tau}_\mu$ are given in Ref.~\cite{Suzuki1987}.

First, we think of the diphoton-dominant case, where $C_L'=2.2\times 10^{-2}$ and $D_L=0$, shown by the red curves.
As indicated by the prefactor of Eq.~\eqref{eq:ene_ang_distribution}, the $Z$-dependence is roughly $Z^5$.
In this case, since the nuclear Coulomb field is necessary to absorb one photon, it is reasonable that the rate disappears in the limit of $Z\to 0$.
Since Eq.~\eqref{eq:ene_ang_distribution} assumes the point-like nuclear charge distribution, it could overestimate the decay rates for large $Z$.
We expect to obtain more realistic calculations by replacing $Z$ with the effective atomic number $Z_\mathrm{eff}$~\cite{Suzuki1987}, which partially includes the finite size of the nucleus.
Applying the correction, we obtain the dotted curves in Fig.~\ref{fig:rate_vs_z_crystalbox}-(a), which show a milder $Z$-dependence.

On the other hand, the blue curves show the dipole-dominant case, where $D_L=1.0\times 10^{-8}$ and $C_L'=0$.
Since the chosen value of $D_L$ is the allowed maximum one, the branching ratio shown in Fig.~\ref{fig:rate_vs_z_crystalbox}-(b) tends to $BR(\mu^+\to e^+\gamma)<3.1\times 10^{-13}$ in the limit of $Z\to 0$.
In contrast to the diphoton operators, the rate slowly becomes smaller as $Z$ is larger.
This is because the decay is affected by the nuclear Coulomb field only through the wave function of the bound muon.
This is analogy to the ordinary muon decay in orbit, where the suppression of large $Z$ is known as the Huff factor~\cite{Huff1961}.
As a result, the $Z$-dependence of the branching ratio is qualitatively different from the case of the diphoton operators; the dipole-dominant case shows decreasing functions of $Z$, while the diphoton-dominant case shows increasing functions.

The muonic atom with higher $Z$ has a shorter lifetime because the nuclear capture rate of a bound muon ($\mu^-p\to \nu_\mu n$) increases with $Z$.
As shown in Fig.~\ref{fig:rate_vs_z_crystalbox}-(b), we find that, as $Z$ increases, $BR\left(\mu^-\to e^-\gamma\right)$ becomes larger for the diphoton case and smaller for the dipole case.
For aluminum with $\tilde{\tau}_\mu=864$\,ns, the maximum allowed branching ratio is $4.6\times 10^{-12}$ with $Z=13$ ($2.7\times 10^{-12}$ with $Z_\mathrm{eff}=11.48$) for the diphoton case, while it is $1.2\times 10^{-13}$ for the dipole case.

We focus on light nuclei in this article; however, our results indicate that heavier nuclei would be more sensitive to diphoton operators.
For heavy nuclei, it will be important to consider the relativistic effects of the bound muon and the distortion of the emitted electron in the nuclear Coulomb field.
More detailed analyses of these effects are left for future work.
To simplify the discussion in the remainder of this article, hereafter we will use $Z$ without the $Z_\mathrm{eff}$ prescription discussed above.

\subsection{Energy and angular distributions of emitted particles}

\begin{figure}[htb]
  \centering
    \begin{tabular}{c}
      \begin{minipage}{0.45\hsize}
		\centering
          \includegraphics[clip, width=7.0cm]{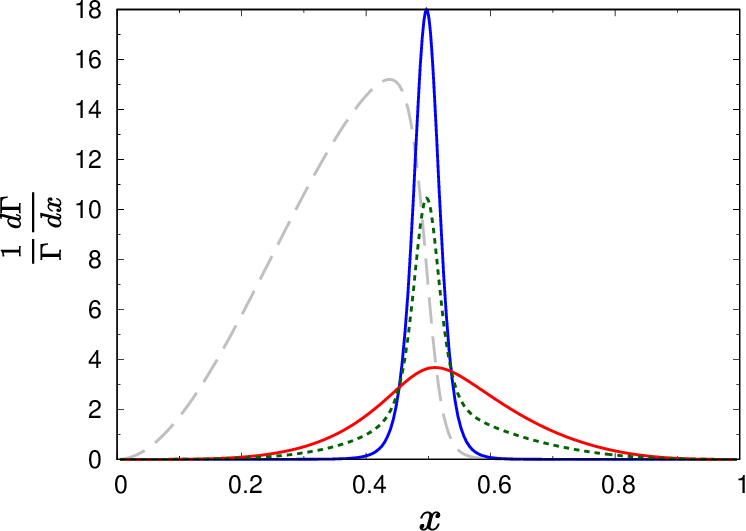}
          \\ (a)
      \end{minipage}%
      \begin{minipage}{0.45\hsize}
        \centering
          \includegraphics[clip, width=7.0cm]{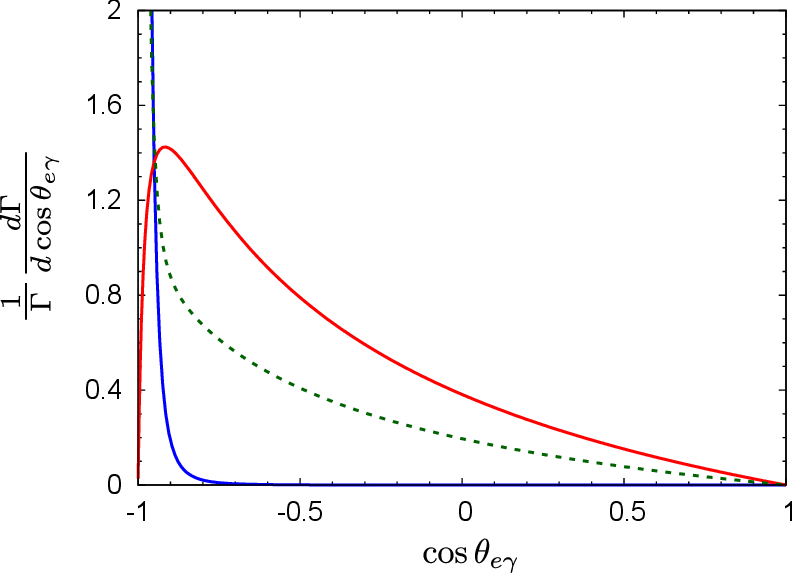}
          \\ (b)
      \end{minipage}%
    \end{tabular}
    \caption{
    (a) The energy distribution of the emitted electron and (b) the angular distribution between the emitted electron and photon (b) for Al with $Z=13$.
   	In (a), the vertical axis is $d\Gamma/dE_e$ divided by $\Gamma$, while the horizontal axis is the dimensionless energy of the electron $x=E_e/m_\mu$.
   	In (b), the vertical axis is $d\Gamma/d\cos\theta_{e\gamma}$ divided by $\Gamma$, while the horizontal axis is the cosine of the opening angle between the electron and photon, $\cos\theta_{e\gamma}$.
   	Red curves indicate $C_L'$ dominant case, while blue curves indicate $D_L$ dominant case.
   	Green dotted curves indicate the case where $D_L:C_L':(C_L+\tilde{C}_L)=2.5\times 10^{-6}:1:1$.
   	In (a), the electron spectrum of the background process ($\mu^-\to e^-\nu_\mu\overline{\nu}_e$) multiplied by 4 is shown by the gray dashed curve for reference.
    }
    \label{fig:dist_Al}
\end{figure}

We calculate the energy and angular distributions of the emitted particles.
In the calculation, an aluminum target is assumed.
For aluminum with $Z=13$, the spectrum of the emitted electrons is given in Fig.~\ref{fig:dist_Al}-(a), and the angular distribution between the emitted electrons and photon is given in Fig.~\ref{fig:dist_Al}-(b).
The diphoton-dominant case is shown by the red curves.
As the back-to-back kinematics of the $\mu^+\to e^+\gamma$ decay in a free muon, the energy of the emitted electron (and also the photon) is approximately given by the half of the muon mass while the opening angle is around $\pi$.
However, the presence of the nuclear Coulomb potential introduces a discrepancy from the approximated picture; the width of the energy spectrum is roughly determined by the Fermi momentum of the bound muon $\sim m_\mu Z\alpha$.

The distributions for the dipole-dominant case are shown by the blue curves.
Compared to the diphoton-dominant case, the dipole-dominant distribution are more similar to back-to-back kinematics, which respectively have sharp peaks at $E_e=E_\mu/2$ and $\cos\theta_{e\gamma}=-1$.
The distributions would help us to discriminate the operators after the discovery of CLFV.

In Fig.~\ref{fig:dist_Al}-(a), the electron spectrum of the background process $\mu^-\to e^-\nu_\mu\overline{\nu}_e$ is shown by the gray dashed curve, which is multiplied by 4 for clarity.
The background spectrum is calculated under the same assumption as the calculation of the $\mu^-\to e^-\gamma$ spectrum.
Unlike the case using free muons for $\mu^+ \to e^+ \gamma$, the peaks of the electron spectrum for the $\mu^-\to e^-\gamma$ signal and the background differ when using bound muons in an atomic orbit.
We expect that the appropriate setting of the energy range would enhance the ability to search for the $\mu^-\to e^-\gamma$ events in the future experiments.

In general, in case the dipole and diphoton contributions are comparable, we would observe the interference of them.
For reference, we also show the case that two operators simultaneously exist by green dotted curves.
In the plot, we set $D_L:C_L':(C_L+\tilde{C}_L)=2.5\times 10^{-6}:1:1$, where the contributions of the dipole and diphoton are comparable and constructively interfere.

We note that the energy spectrum of the photon is immediately obtained by reversing that of the electron because the photon energy is given by $E_\gamma=E_\mu-E_e$ neglecting the nuclear recoil energy.

\subsection{Invariant mass distribution}

In addition to $E_e$ and $\cos\theta_{e\gamma}$, we also focus on the invariant mass of the emitted electron and photon,
\begin{align}
m_{e\gamma}=\sqrt{\left(E_e+E_\gamma\right)^2-\left|\bm{p}_e+\bm{p}_\gamma\right|^2}=m_\mu\sqrt{2xyd+r},
\end{align}
which is allowed in the range of $m_e\le m_{e\gamma}\le E_\mu$.
In searches for the free muon decay $\mu^+\to e^+\gamma$, the invariant mass $m_{e\gamma}$ is an important variable because $m_{e\gamma}=m_\mu$ is satisfied for a signal.
Although the $\mu^-\to e^-\gamma$ process in a muonic atom does not respect the strict relation, the kinematical cutoff for the invariant mass is also useful to reduce the accidental coincidence backgrounds.

\begin{figure}[htb]
  \centering
  \includegraphics[clip, width=7.0cm]{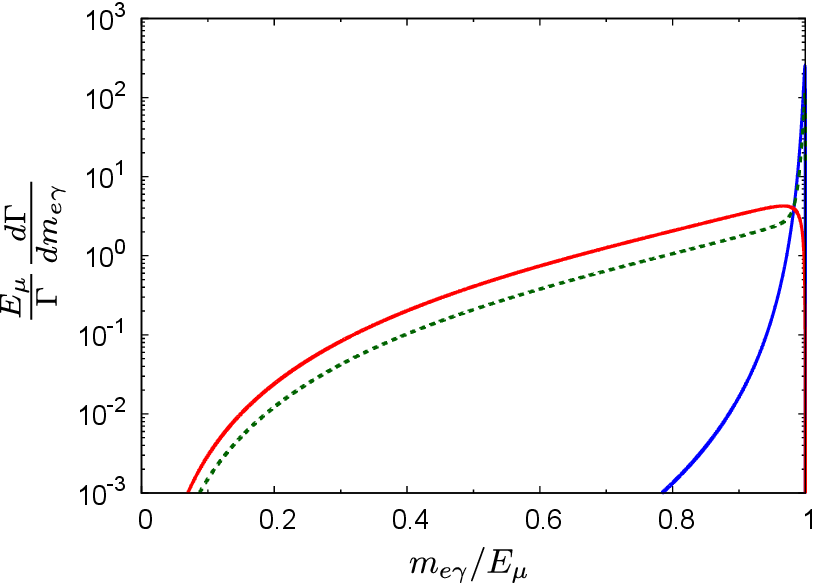}
  \caption{
  The distribution of the invariant mass $m_{e\gamma}$ between the emitted electron and photon for Al with $Z=13$.
  The vertical axis is $d\Gamma/dm_{e\gamma}$ times $E_\mu/\Gamma$, while the horizontal axis is $m_{e\gamma}$ divided by $E_\mu$.
  See the caption of Fig.~\ref{fig:dist_Al} for the assumed parameters of each curve.
  }
  \label{fig:dist_invmass_Al}
\end{figure}

Based on Eq.~\eqref{eq:ene_ang_distribution}, the $m_{e\gamma}$ distribution can be calculated by
\begin{align}
\frac{d\Gamma}{dm_{e\gamma}}=\frac{m_{e\gamma}}{m_\mu^2}\int_{E_e^-/m_\mu}^{E_e^+/m_\mu}\frac{dx}{y\sqrt{x^2-r}}\int_{-1}^{1}d\cos\theta_{e\gamma}\frac{d\Gamma}{dxd\cos\theta_{e\gamma}}\delta\left(\cos\theta_{e\gamma}-\frac{E_\mu^2-p_e^2-E_\gamma^2-m_{e\gamma}^2}{2p_eE_\gamma}\right),
\end{align}
where the upper and lower values of the integral variable $E_e$ are given as
\begin{align}
E_e^\pm=\frac{E_\mu}{2}\left(1+\frac{m_e^2}{m_{e\gamma}^2}\right)\pm\frac{\sqrt{E_\mu^2-m_{e\gamma}^2}}{2}\left(1-\frac{m_e^2}{m_{e\gamma}^2}\right).
\end{align}
The numerical result is shown by Fig.~\ref{fig:dist_invmass_Al} for the case of aluminum with $Z=13$.
As seen in the figure, the distribution is typically located to $m_{e\gamma}\simeq E_\mu$, especially for the dipole dominant case shown by the blue curve.
The cutoff for low $m_{e\gamma}$ would work to reduce backgrounds in experimental searches.

\section{Signal sensitivity and background estimates \label{sec:Background}}

We will evaluate the signal sensitivity for $\mu^{-}\to e^{-}\gamma$ in a muonic atom and examine the corresponding backgrounds. There are two major backgrounds to the search for $\mu^{-}\rightarrow e^{-} \gamma$, similar to those in the search for $\mu^{+}\rightarrow e^{+} \gamma$ decay at rest. The first is a physics background from radiative muon decay, $\mu^-\to e^-\nu_\mu\overline{\nu}_e\gamma$, where the electron and photon are emitted back-to-back, with the two neutrinos carrying away a small amount of energy. The second is an accidental coincidence background, where an electron from a normal muon decay, $\mu^-\to e^-\nu_\mu\overline{\nu}_e$, coincides with a high-energy photon. One potential source of the high-energy photon could be another muon decaying as $\mu^-\to e^-\nu_\mu\overline{\nu}_e\gamma$. It is important to note that while there may be potential background contributions from the products of nuclear muon capture in a muonic atom, this study does not address them.

In rare decay searches, it is crucial to identify signal events while minimizing background contribution, particularly for $\mu^{-} \to e^{-} \gamma$, as the emitted electron and photon in this decay have non-monochromatic spectra. For simplicity, the signal event selection in this preliminary study includes three requirements: energy sum conservation between the electron and photon, time coincidence between the two, and angular constraints reflecting near back-to-back kinematics.

It should be noted that in the experimental search for $\mu^{+}\to e^{+}\gamma$ decay, the energy selection criteria are stricter, requiring the electron and photon energies to be $m_{\mu}/2$ individually, rather than focusing on the energy sum. While comprehensive event selection criteria or optimization of the signal event selection could be applied to $\mu^{-}\to e^{-}\gamma$, these are not implemented in this study, as they are beyond the scope of our current investigation.

In current and future experiments, the increasing flux of incident muons causes the accidental coincidence background to dominate over the physics background. 
To proceed with further investigation, we define an effective branching ratio as the estimated event rate after the signal event selection, normalized to the total decay rate of a muonic atom ($1/\tilde{\tau}_{\mu}$). The effective branching ratio for the accidental coincidence background, $B_{\rm acc}$, can be expressed as follows~\cite{Kuno:1999jp}:
\begin{equation}
B_{\rm acc} = R_{\mu}\cdot \Delta t_{e\gamma} \cdot {\Delta\Omega_{e\gamma} \over 4\pi}\cdot f_{acc},
\end{equation}
\noindent where $R_{\mu}$ is the number of muons available for the search per unit time. 
$\Delta t_{e\gamma}$ and $\Delta\Omega_{e\gamma}$ represent the total widths of the signal regions for timing coincidence and angular constraints of the back-to-back kinematics, respectively.
The factor $f_{acc}$ represents the probability that an electron emitted from one muon and a photon emitted from another muon satisfy the requirement of energy sum conservation ($x+y$) where the total width of the signal region is given by $\Delta_{x+y}$;
\begin{align}
    f_{acc}=&\tilde{\tau}_\mu\int dx\frac{d\Gamma\left(\mu^-\to e^-\nu_\mu\overline{\nu}_e\right)}{dx}\cdot\tilde{\tau}_\mu\int dy\frac{d\Gamma\left(\mu^-\to e^-\nu_\mu\overline{\nu}_e\gamma\right)}{dy}\cdot\theta\left(\Delta_{x+y}-|1-b-x-y|\right).
\end{align}
The explicit formulas of the differential decay rates of $\mu^{-}\to e^{-}\nu_{\mu}\overline{\nu}_{e}$ and $\mu^{-}\to e^{-}\nu_{\mu}\overline{\nu}_{e}\gamma$ decays are provided in Appendix~\ref{app:electron_BG} and \ref{app:photon_BG}, respectively.

\begin{figure}[htb]
  \centering
  \includegraphics[clip, width=7.0cm]{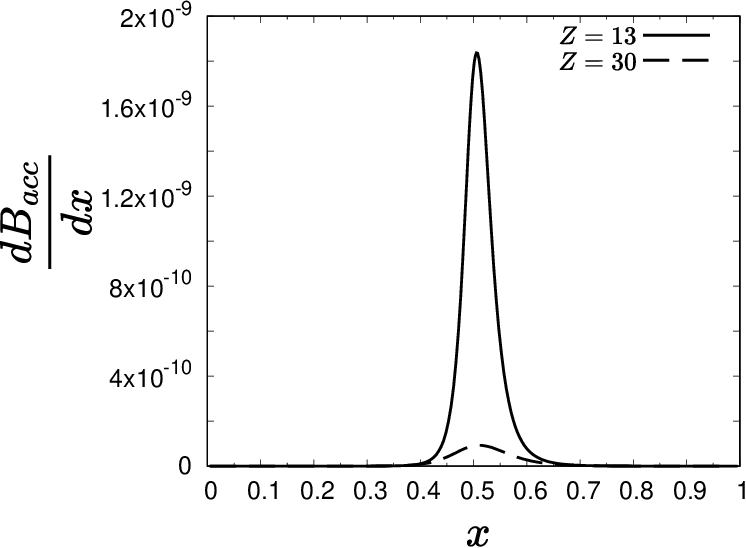}
  \caption{
  The energy distribution of electrons for the accidental coincidence background with the energy sum requirement.
  The solid and dashed lines correspond to aluminum ($Z=13$) and zinc ($Z=30$) targets, respectively.
  We use $R_\mu=10^7$/s, $\Delta t_{e\gamma}=78$\,ps and $\Delta_{x+y}=0.01$.
  }
  \label{fig:x_dist_acc_bg}
\end{figure}

Given the sizes of the signal regions, one can evaluate the effective branching ratios of both the signals and the backgrounds. In our estimation, we use $R_\mu=10^7$/s, $\Delta t_{e\gamma}=78$\,ps and $\Delta_{x+y}=0.01$. The values of $\Delta t_{e\gamma}$ and $\Delta_{x+y}$ are chosen based on the detector resolutions achieved in the MEG II experiment~\cite{MEGII:2023ltw}. It is noted that $\Delta_{x+y}$ is determined by the relatively poorer photon energy resolution, which was reported to be approximately 1.8\,\% at photon energy of 52.8~MeV, leading $\Delta_{x+y} = 0.01$~\cite{MEGII:2023ltw}. The energy spectrum of the accidental coincidence background is shown in Fig.~\ref{fig:x_dist_acc_bg}, where the angular constraint of $\Delta\Omega_{e\gamma}/4\pi$ is not applied. Note that, due to the requirement of the energy sum conservation, the spectrum is peaked around $x=(1-b)/2$, which closely resembles the signal spectrum. This indicates that the energy distribution alone may be less effective for distinguishing the signal from the accidental coincidence background, after the energy sum requirement.

To assess whether the search can be conducted without significant background contributions, we estimate the effective branching ratios for both the signals and the backgrounds, including the accidental coincidence and physics backgrounds, using the signal selection criteria mentioned above, initially without the angular constraint of $\Delta\Omega_{e\gamma}$.
They are shown in the second and third columns of Table~\ref{tab:total_effective_branching_ratio}. The results are presented for aluminum ($Z=13$) and zinc ($Z=30$). In the estimation for the diphoton-dominant (dipole-dominant) signal, we set $C_L'=2.2\times 10^{-2}$ ($D_L=1.0\times 10^{-8}$) with the other couplings set to zero.
When comparing the results, we find that the accidental coincidence background significantly exceeds the CLFV signals in both the dipole-dominant and diphoton-dominant cases, while the physics background from radiative muon decay remains negligible compared to the current constraints on the signals.
\begin{table}[htb]
\centering
 \begin{tabular}{l|c|c|c|c}
  & \multicolumn{2}{c}{without $\theta_{e\gamma}$ constraint} & \multicolumn{2}{|c}{with $\theta_{e\gamma}$ constraint} \\
  \cline{2-5}
  & Aluminum ($Z=13$) & Zinc ($Z=30$) & Aluminum ($Z=13$) & Zinc ($Z=30$) \\
 \hline\hline
Signal (diphoton-dominant) & $4.6\times 10^{-12}$ & $2.2\times 10^{-11}$ & $7.1\times 10^{-14}$ & $3.3\times 10^{-13}$ \\
Signal (dipole-dominant) & $1.2\times 10^{-13}$ & $2.0\times 10^{-14}$ & $1.2\times 10^{-17}$ & $1.6\times 10^{-17}$ \\
Accidental coincidence background & $1.2\times 10^{-10}$ & $1.2\times 10^{-11}$ & $8.5\times 10^{-13}$ & $1.1\times 10^{-13}$ \\
Physics background & $1.5\times 10^{-15}$ & $5.7\times 10^{-16}$ & $2.6\times 10^{-19}$ & $6.1\times 10^{-19}$
\end{tabular}
\caption{Effective branching ratios of signal and backgrounds for aluminum ($Z=13$) and zinc ($Z=30$).
The second and third columns show the cases without the angular constraint.
The fourth and fifth columns show the cases with the angular constraint, Eq.~\eqref{eq:angular_cut}.}
\label{tab:total_effective_branching_ratio}
\end{table}

To further reduce the accidental coincidence background, we consider applying the angular constraint of $\Delta\Omega_{e\gamma}$.
As shown in Fig.~\ref{fig:dist_Al}-(b), the angular distribution of the signal process is concentrated around $\theta_{e\gamma} = \pi$, while the distribution for the accidental coincidence background is flat.
Therefore, the angular constraint is expected to be effective in distinguishing the signal from the background.

We will consider optimizing the angular constraint of $\Delta\Omega_{e\gamma}$ for the diphoton-dominant signal. The $\cos\theta_{e\gamma}$ distribution of the diphoton-dominant signal is centered around $\cos\theta_{e\gamma} = -1$, although it drops to zero exactly at $\cos\theta_{e\gamma} = -1$.
Therefore, we apply the following angular constraint:
\begin{align}
    \theta_{e\gamma}^{\rm peak}-\frac{\Delta\theta_{e\gamma}}{2}<\theta_{e\gamma}<\theta_{e\gamma}^{\rm peak}+\frac{\Delta\theta_{e\gamma}}{2},
    \label{eq:angular_cut}
\end{align}
with $\Delta\theta_{e\gamma}=0.02$ rad. This value of $\Delta\theta_{e\gamma}$ was also adopted, as the angular resolution observed in the MEG II/MEG experiments was reported to be about 17\,mrad~\cite{MEGII:2023ltw,MEGdetector2013}. The angle $\theta_{e\gamma}^{\rm peak}$ is the peak value of the $\theta_{e\gamma}$ distribution: $d\Gamma/d\theta_{e\gamma}=\left(\sin\theta_{e\gamma}\right)d\Gamma/d\cos\theta_{e\gamma}$ is maximized when $\theta_{e\gamma}=\theta_{e\gamma}^{\rm peak}$.
We numerically find $\theta_{e\gamma}^{\rm peak}=2.39$ rad for aluminum and $\theta_{e\gamma}^{\rm peak}=2.13$ rad for zinc. By applying the angular constraint, we obtain the effective branching ratios shown in the fourth and fifth columns of Table~\ref{tab:total_effective_branching_ratio}.

Compared to the values without the angular constraint, we find that the angular constraint reduces the accidental coincidence background by a factor of approximately $\Delta \Omega_{e\gamma}/4\pi \simeq \left(\sin\theta_{e\gamma}^{\rm peak}\right)\Delta\theta_{e\gamma}/2$. Meanwhile, the diphoton-dominant signal largely preserves its signal acceptance. For zinc, the effective branching ratio for the maximum allowed signal exceeds the estimation of the accidental coincidence background by a factor of about three, as the signal rate increases with larger $Z$ in the diphoton-dominant case, even though the fraction of muon decay in an atomic orbit is reduced with $\tilde{\tau}_{\mu}$. 

This suggests that future planned experiments could improve the limit by a factor of more than three for targets heavier than zinc. It is important to note that this study provides an initial estimation and does not account for detailed detector parameters, such as geometrical acceptance. Further optimization can be explored in future experimental planning. In particular, the accidental coincidence background depends on the muon beam rate, $R_{\mu}$. Therefore, if $R_{\mu}$ is reduced from its reference rate, the background contribution from accidental coincidences can be reduced relative to the signal rate, allowing for improved experimental sensitivity. However, this reduction would require the experiment to run for a longer period and necessitate further optimization of the experimental design.

Another interesting decay, $\mu^{-} \to e^{-} a$, where $a$ could be an axion-like particle, followed by $a \to \gamma\gamma$ with one of the photons absorbed by the nucleus, has a similar event signature to the one in the current study. We will explore this decay mode in our upcoming publication.

\section{Summary \label{sec:Summary}}

Our study explores the potential of the new CLFV process $\mu^-\to e^-\gamma$ in a muonic atom. This CLFV process shows promise for direct sensitivity to the diphoton effective CLFV operator. Our work includes the derivation of formulas for the decay rates and kinematic distributions of $\mu^{-}\to e^{-}\gamma$, along with $\mu^{-}\to e^{-}\nu_{\mu}\overline{\nu}_{e}$ and $\mu^{-}\to e^{-}\nu_{\mu}\overline{\nu}_{e}\gamma$, providing valuable information for future investigations in CLFV searches in a muonic atom.

Based on the constraints from the Crystal Box experiment~\cite{CrystalBox}, we estimate that the search for $\mu^-\to e^-\gamma$ could be improved for a muonic atom of a heavy element such as zinc or heavier targets, while maintaining a lower background level.

Although we investigate for light nuclei in this article, our results show that heavy nuclei would be more sensitive to the diphoton operators.
For the heavy nuclei, more detailed calculation would be necessarily to take into account the relativistic effect of muon and the Coulomb distortion of the emitted electrons.
We leave the analyses for heavy nuclei in future work.

\begin{acknowledgments}
We thank S.~Davidson for her collaboration in the initial stage of this work.
YU thanks Profs.~J.~Sato, Y.~Kiyo, and K.~Ohkuma for fruitful discussions.
YK thanks the Department of Physics at University of Science and Technology of China for their warm hospitality during his visiting stay.
This work was supported by the Japan Society for the Promotion of Science (JSPS) KAKENHI, under Grant Numbers JP22K03602, JP23K13106, and JP23K22508 (YU), JP18H05231 (YK), and 22K03638, 22K03602, JP20H05852 (MY).
This work was partly supported by MEXT Joint Usage/Research Center on Mathematics and Theoretical Physics JPMXP0619217849 (MY).
\end{acknowledgments}

\appendix

\section{Background electrons in muonic atoms \label{app:electron_BG}}

The explicit formula for the electron spectrum of muon decay, $\mu^- \to e^- \nu_\mu \overline{\nu}_e$, in a muonic atom is discussed. The quantitative formula, even for heavy nuclei, is provided in Refs.\cite{Watanabe:1987su,Czarnecki:2011mx}, and the radiative corrections have been considered in Refs.\cite{Szafron:2015kja,Szafron:2016cbv}. For the purposes of this article, it is sufficient to use the formula valid for light nuclei; we neglect the relativistic effects of the muon wave functions, the Coulomb distortion of the electron wave functions, and the finite size of the nucleus.

The simple formula for the spectrum of the decay in orbit is derived as
\begin{align}
    &\frac{1}{\Gamma_0}\frac{d\Gamma\left(\mu^-\to e^-\nu_\mu\overline{\nu}_e\right)}{dx} \nonumber\\
    =&\frac{512}{3\pi}\zeta^5\sqrt{x^2-r}\int_{0}^{k_0}dkk^2\frac{x\left(3k_0^2-k^2\right)\left\{3\left(k^2+x^2-r+\zeta^2\right)^2+4k^2\left(x^2-r\right)\right\}+16k_0k^2\left(x^2-r\right)\left(k^2+x^2-r+\zeta^2\right)}{\left\{k^4-2\left(x^2-r-\zeta^2\right)k^2+\left(x^2-r+\zeta^2\right)^2\right\}^3},
\end{align}
where $k_0=1-b-x$.
In the limit of $Z\to 0$, this formula reproduces the decay spectrum of the free muon,
\begin{align}
    \frac{1}{\Gamma_0}\frac{d\Gamma\left(\mu^+\to e^+\overline{\nu}_\mu\nu_e\right)}{dx}=& 16\sqrt{x^2-r}\left\{3x-4x^2-(2-3x)r\right\}\theta\left(\frac{1+r}{2}-x\right),
\end{align}
where $\theta(z)$ is the step function satisfying $\theta(z)=1$ if $z\ge 0$ and $\theta(z)=0$ if $z<0$.

\begin{figure}[htb]
  \centering
  \includegraphics[clip, width=9.0cm]{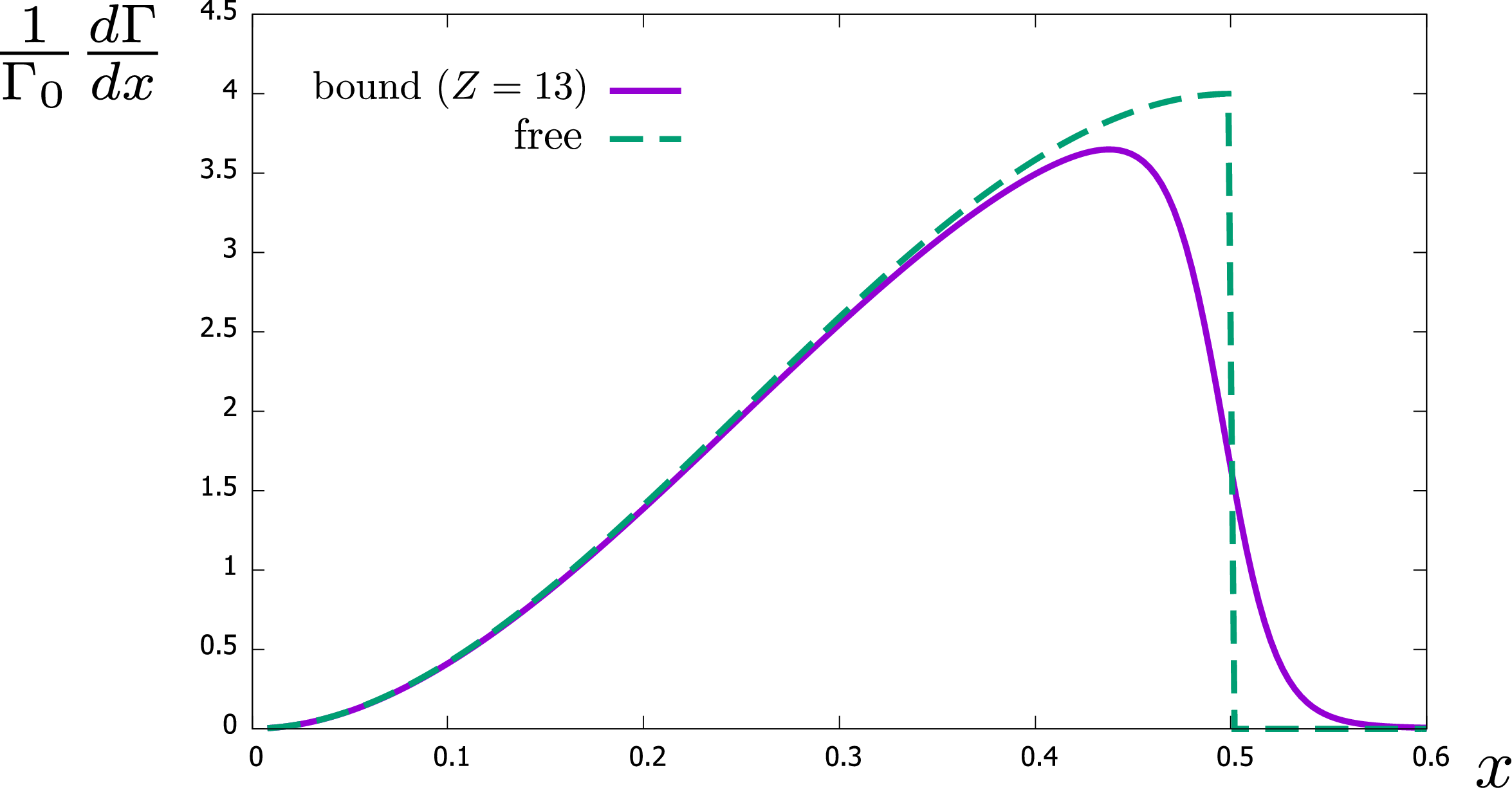}
  \caption{
  The energy distribution of the emitted electron in the $\mu^+\to e^+\overline{\nu}_\mu\nu_e$ process for free muons and the $\mu^-\to e^-\nu_\mu\overline{\nu}_e$ process for bound muons.
  The solid curve is the case of the bound muon decay for Al with $Z=13$, and the dashed curve is the case of the free muon decay.
  }
  \label{fig:DIO}
\end{figure}

The solid curve of Fig.~\ref{fig:DIO} shows the $x$-distribution for $Z=13$.
The dashed curve shows the distribution for the free muon decay, where the endpoint energy is given by $(1+r)/2$.
The free muon spectrum has a steep edge at the endpoint, while the bound muon spectrum is smeared and the edge gets mild.

\section{Background photons in muonic atoms \label{app:photon_BG}}

The explicit formula for the photon spectrum of radiative decay of the bound muon, $\mu^- \to e^- \nu_\mu \overline{\nu}_e \gamma$, in a muonic atom is presented. In our analysis, the relativistic effects of the muon wave functions, the Coulomb distortion of the electron wave functions, and the finite size of the nucleus are neglected.
The differential decay rate is given by
\begin{align}
    \frac{1}{\Gamma_0}\frac{d\Gamma\left(\mu^-\to e^-\nu_\mu\overline{\nu}_e\gamma\right)}{dxdyd\cos\theta_{e\gamma}}=&\frac{\alpha}{4\pi}\frac{\sqrt{x^2-r}}{x}\frac{1}{y}F(x,y,d;Z), \\    F(x,y,d;Z)=&\int_{0}^{k_0}dkk^2\int_{-1}^{1}dc\frac{16}{\pi}\frac{\zeta^5}{\left(w^2+k^2+2kwc+\zeta^2\right)^{4}} \nonumber\\
    &\times 16\left[F_{00}\left(k_0,k,x,y,d\right)+\frac{y}{Y\left(k_0,k,c\right)}F_{01}\left(k_0,k,x,y,d\right)+\left(\frac{y}{Y\left(k_0,k,c\right)}\right)^3F_{03}\left(k_0,k,x,y,d\right)\right. \nonumber\\
    &+\left\{F_{10}\left(k_0,k,x,y,d\right)+\frac{y}{Y\left(k_0,k,c\right)}F_{11}\left(k_0,k,x,y,d\right)+\left(\frac{y}{Y\left(k_0,k,c\right)}\right)^3F_{13}\left(k_0,k,x,y,d\right)\right\}c \nonumber\\
    &\left.+\left\{\frac{y}{Y\left(k_0,k,c\right)}F_{21}\left(k_0,k,x,y,d\right)+\left(\frac{y}{Y\left(k_0,k,c\right)}\right)^3F_{23}\left(k_0,k,x,y,d\right)\right\}c^2\right],
    \label{eq:bound_radiative}
\end{align}
where $k_0=1-b-x-y$.
Here, we define
\begin{align}
    Y\left(k_0,k,c\right)=&\frac{1}{2w}\sqrt{\frac{y^2\left\{u^2-4k^2\left(x^2-r\right)\right\}\left\{x^2d\left(2-d\right)-r\right\}+\left[\left\{x^2-r+xy(1-d)\right\}u+2kw\left(x^2-r\right)c\right]^2}{x^2-r}},
\end{align}
with $u=1+x^2-r+k^2-\left(k_0+x\right)^2$.
We note that $Y\left(1-x-y,w,-1\right)=y$ is satisfied.
The explicit formulas of $F_{00}$, $F_{01}$, $F_{03}$, $F_{10}$, $F_{11}$, $F_{13}$, $F_{21}$, and $F_{22}$ are given as follows:
\begin{align}
    F_{00}\left(k_0,k,x,y,d\right)=&\frac{2}{d^2x}\left(3k_0^2-k^2\right)\left[(x+y)\left\{dx^2(2-d)-r\right\}+dxy^2\right] \nonumber\\
    &-\frac{2}{d}h_1-\frac{u}{d}\left\{(1-2x)(x+y)+r\right\}+2x\left\{u(1-x-y)-2y^2(1-x)\right\}, \\
    F_{01}\left(k_0,k,x,y,d\right)=& \frac{u}{dy}h_1-\frac{2k_0^2}{d}\left[dx\left\{x+y+x(1+2y)(1-d)+2(2-d)x^2\right\}-r(1+2x+y)\right] \nonumber\\
    &+\frac{2k_0^3}{d}\left[dx\left\{(2-d)x+y\right\}-r\right]+\frac{2k^2}{d}\left[dx\left\{y\left(2-k_0\right)+x\left(2-d-2k_0+dk_0-2dy\right)\right\}-\left(1-k_0\right)r\right] \nonumber\\
    &-\frac{u^2}{2dy}\left\{2dx(1-x-y)-(1-2x)(x+y)-r\right\}, \\
    F_{03}\left(k_0,k,x,y,d\right)=&\frac{x}{2y}\left[h_2u+u^3(1-x)-\frac{4k^2y^2}{w^2}\left\{d(2-d)x^2-r\right\}\left\{h_3+2u(1-x)\right\}\right], \\
    F_{10}\left(k_0,k,x,y,d\right)=&-\frac{2kw}{d^2x}\left[\frac{dx}{w^2}\left\{2k_0x(1-d)+2dx(1-x-y)-(1-2x)(x+y)-r\right\}\left\{x^2-r+xy(1-d)\right\}\right. \nonumber\\
    &\left.+2k_0\left\{xd(x+y)-r\right\}-dx\left\{y+2dx(1-x-y)+2x^2-2r\right\}\right], \\
    F_{11}(k_0,k,x,y,d)=&-\frac{kw}{dy}\left[\left\{y+2dx(1-x-y)+2x^2-2r\right\}u\right. \nonumber\\
    &\left.+2k_0\left\{x(1-d)\left(2x^2-y\right)-2x^2\left(1-d+dy\right)+2ry\right\}-2\left(k_0^2-k^2\right)x\left\{x-d\left(x+y\right)\right\}\right], \\
    F_{13}\left(k_0,k,x,y,d\right)=&\frac{kx}{wy}\left\{x^2-r+xy(1-d)\right\}\left\{h_2+uh_3+3u^2(1-x)\right\}, \\
    F_{21}\left(k_0,k,x,y,d\right)=&-\frac{2k^2}{dy}w^2(x-r), \\
    F_{23}\left(k_0,k,x,y,d\right)=&\frac{2k^2}{y}x\left(x^2-r\right)\left\{h_3+2u(1-x)\right\},
\end{align}
where
\begin{align}
    h_1=&\left(k_0^2-k^2\right)(1-d)x(x+y) \nonumber\\
    &+k_0\left[y^2+3xy+2x^2(1-y)-2x^3-dx\left\{y(1+2y)+2x(1+y)-2x^2\right\}+2d^2x^2y\right], \\
    h_2=& k_0^2\left\{x\left(3-4x+4x^2\right)-r(2+x)\right\}-2k_0^3\left\{x(3-2x)-r\right\}+3k_0^4x \nonumber\\
    &-k^2\left[x\left(1+2k_0^2\right)-2k_0\left\{x(3-2x)-r\right\}-r(2-x)\right]-k^4x, \\
    h_3=& -k_0\left\{(1-2x)^2-r\right\}+k_0^2(3-2x)-2k_0^3-k^2\left(3-2x-2k_0\right).
\end{align}

For reference, the differential decay rate of the free muon decay, $\mu^+\to e^+\overline{\nu}_\mu\nu_e\gamma$, is given by \cite{Fronsdal:1959zzb}
\begin{align}
    \frac{1}{\Gamma_0}\frac{d\Gamma\left(\mu^+\to e^+\overline{\nu}_\mu\nu_e\gamma\right)}{dxdyd\cos\theta_{e\gamma}}=&\frac{\alpha}{4\pi}\frac{\sqrt{x^2-r}}{x}\frac{1}{y}F(x,y,d), \\
    F(x,y,d)=& F^{(0)}(x,y,d)+rF^{(1)}(x,y,d)+r^2F^{(2)}(x,y,d),
    \label{eq:free_radiative}
\end{align}
where
\begin{align}
    F^{(0)}(x,y,d)=&\frac{32}{d}\left\{y^2(3-4y)+6xy(1-2y)+2x^2(3-8y)-8x^3\right\} \nonumber\\
    &-32x\left\{y(3-2y-4y^2)+x(3-2y-16y^2)-4x^2(1+4y)\right\} \nonumber\\
    &+32dx^2y\left\{3-5y-4y^2-4x(2+3y)\right\}+128d^2x^3y^2(1+y), \\
    F^{(1)}(x,y,d)=&\frac{32}{d^2}\left\{-\frac{y(3-4y)}{x}-3+8y+4x\right\}+\frac{32}{d}\left\{y(3-5y)-2x(2+y)+6x^2\right\} \nonumber\\
    &+32x\left\{2-3y+2y^2-3x(1+2y)\right\}+96dx^2y(1+y), \\
    F^{(2)}(x,y,d)=&\frac{32}{d^2}\left(\frac{2-3y}{x}-3\right)+\frac{96}{d}y.
\end{align}

In the limit $Z\to 0$, Eq.~\eqref{eq:bound_radiative} must agree with Eq.~\eqref{eq:free_radiative}.
For an arbitrary function $f(k,c)$, we find
\begin{align}
    \lim_{Z\to 0}\int_{0}^{k_0}dkk^2\int_{-1}^{1}dc\frac{16}{\pi}\frac{\zeta^5}{\left(k^2+w^2+2kwc+\zeta^2\right)^{4}}f(k,c)=f(w,-1)\theta\left(1-x-y-w\right).
\end{align}
Using the relation, the limit of $Z\to 0$ gives us the expected result,
\begin{align}
    \lim_{Z\to 0}F(x,y,d;Z)=F(x,y,d).
\end{align}
We note that the allowed region of the kinematical variables $(x,y,\cos\theta_{e\gamma})$ is restricted for the $Z\to 0$ limit, which can be represented by the step function $\theta\left(1-x-y-w\right)$.

\begin{figure}[htb]
  \centering
  \includegraphics[clip, width=9.0cm]{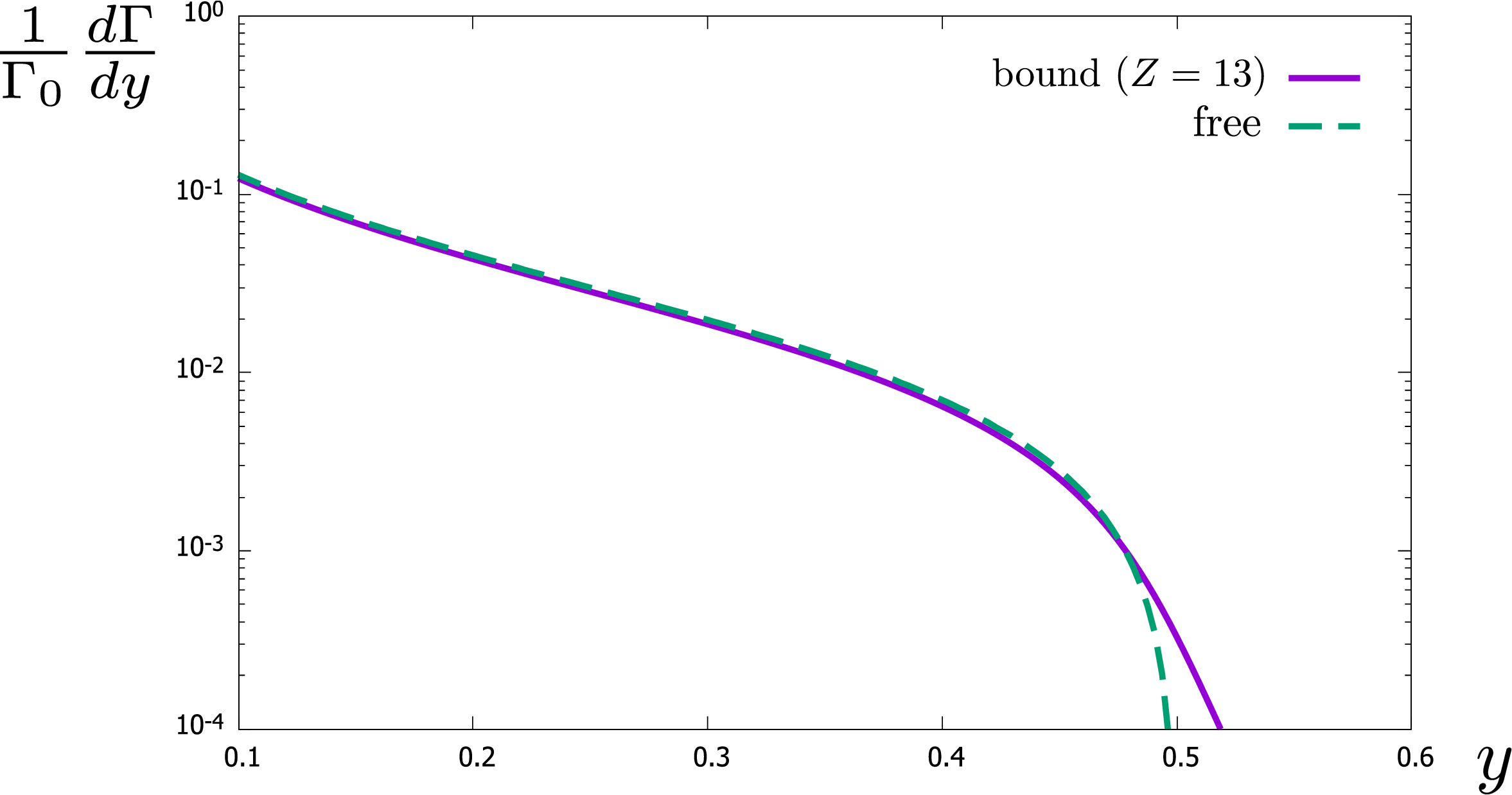}
  \caption{
  The energy distribution of the emitted photon in the $\mu^+\to e^+\overline{\nu}_\mu\nu_e\gamma$ process for free muons and the $\mu^-\to e^-\nu_\mu\overline{\nu}_e\gamma$ process for bound muons.
  The solid curve is the case of the bound muon decay for Al with $Z=13$, and the dashed curve is the case of the free muon decay.
  The vertical axis is displayed in logarithmic scale.
  }
  \label{fig:radiative}
\end{figure}

Figure~\ref{fig:radiative} shows the $y$-distribution.

\end{document}